\documentclass[sigconf]{acmart}

\usepackage{soul}
\usepackage{url}
\usepackage{graphicx}
\usepackage{amsmath}
\usepackage{amsthm}
\usepackage{subfigure}
\usepackage{booktabs}
\usepackage{enumitem}
\usepackage{bm}
\usepackage{xspace}
\usepackage{multirow}
\usepackage{subcaption}
\usepackage[linesnumbered,ruled,vlined]{algorithm2e}
\usepackage{stfloats}

\newtheorem{definition}{Definition}
\newcommand{\model}{\textsc{CoTrans}}
\newtheorem{lemma}{Lemma}

\newtheorem{myremark}{Remark}

\AtBeginDocument{%
 }

\copyrightyear{2024}
\acmYear{2024}
\setcopyright{rightsretained}
\acmConference[CIKM '24]{Proceedings of the 33rd ACM International Conference on Information and Knowledge Management}{October 21--25, 2024}{Boise, ID, USA}
\acmBooktitle{Proceedings of the 33rd ACM International Conference on Information and Knowledge Management (CIKM '24), October 21--25, 2024, Boise, ID, USA}
\acmDOI{10.1145/3627673.3679595}
\acmISBN{979-8-4007-0436-9/24/10}

\begin{document}

\title{The Devil is in the Sources! Knowledge Enhanced Cross-Domain \\ Recommendation in an Information Bottleneck Perspective}

\author{Binbin Hu}
\authornote{Both authors contributed equally to this research.}
\email{bin.hbb@antfin.com}
\affiliation{%
  \institution{Ant Group}
  \country{China}
  \city{Hangzhou}
}

\author{Weifan Wang}
\authornotemark[1]
\email{weifan.wwf@antgroup.com}
\affiliation{%
  \institution{Ant Group}
  \country{China}
  \city{Hangzhou}
}

\author{Shuhan Wang}
\email{hanshu.wsh@antgroup.com	}
\affiliation{%
  \institution{Ant Group}
  \country{China}
  \city{Hangzhou}
}

\author{Ziqi Liu}
\email{ziqiliu@antgroup.com	}
\affiliation{%
  \institution{Ant Group}
  \country{China}
  \city{Hangzhou}
}

\author{Bin Shen}
\email{ringo.sb@antgroup.com}
\affiliation{%
  \institution{Ant Group}
  \country{China}
  \city{Hangzhou}
}

\author{Yong He}
\email{heyong.h@antgroup.com}
\affiliation{%
  \institution{Ant Group}
  \country{China}
  \city{Hangzhou}
}

\author{Jiawei Chen}
\authornote{Corresponding author.}
\email{sleepyhunt@zju.edu.cn}
\affiliation{%
  \institution{Zhejiang University}
  \country{China}
  \city{Hangzhou}
}

\renewcommand{\shortauthors}{Binbin Hu et al.}
\begin{abstract}

Cross-domain Recommendation (CDR) aims to alleviate the data sparsity and the cold-start problems in traditional recommender systems by leveraging knowledge from an informative source domain.
However, previously proposed CDR models pursue an imprudent assumption that the entire information from the source domain is equally contributed to the target domain, neglecting the evil part that is completely irrelevant to users' intrinsic interest.
To address this concern, in this paper, we propose a novel knowledge enhanced cross-domain recommendation framework named {\textsc{CoTrans}}, which remolds the core procedures of CDR models with: 
\textit{Compression} on the knowledge from the source domain and \textit{Transfer} of the purity to the target domain.
Specifically, following the theory of Graph Information Bottleneck, {\textsc{CoTrans}} first compresses the source behaviors with the perception of information from the target domain.
Then to preserve all the important information for the CDR task, the feedback signals from both domains are utilized to promote the effectiveness of the transfer procedure.
Additionally, a knowledge-enhanced encoder is employed to narrow gaps caused by the non-overlapped items across separate domains.
Comprehensive experiments on three widely used cross-domain datasets demonstrate that {\textsc{CoTrans}} significantly outperforms both single-domain and state-of-the-art cross-domain recommendation approaches. 
\end{abstract}







\newcommand{\paratitle}[1]{\vspace{1.5ex}\noindent\textbf{#1}}
\newcommand{\ie}{\emph{i.e.,}\xspace}
\newcommand{\aka}{\emph{a.k.a.,}\xspace}
\newcommand{\eg}{\emph{e.g.,}\xspace}
\newcommand{\etal}{\emph{et al.}\xspace}
\newcommand{\wrt}{\emph{w.r.t.}\xspace}

\newcommand{\ignore}[1]{}
\newcommand{\note}[1]{{\textcolor{red}{[#1]}}}
\newcommand{\tian}[1]{{\textcolor{black}{#1}}}

\maketitle

\section{Introduction\label{sec:Intro}}
To alleviate information overload on the web and understand user preferences, recommender systems have become the key component for online marketplaces and achieved great success.
Over the past decades, a variety of strategies have been developed to harness users' historical behaviors to enhance the quality of recommendation, ranging from traditional collaborative filtering~\cite{linden2003amazon,sarwar2001item} and deep neural networks~\cite{he2017neural} to sequential modeling~\cite{lei2021semi} and graph learning~\cite{liu2021tail,guo2021dual}.
Despite the advancements, recommender systems still grapple with the issues of data sparsity and the cold-start problem, significantly hindering the ability to capture user interests accurately and efficiently.

\begin{figure}
  \includegraphics[width=0.47\textwidth]{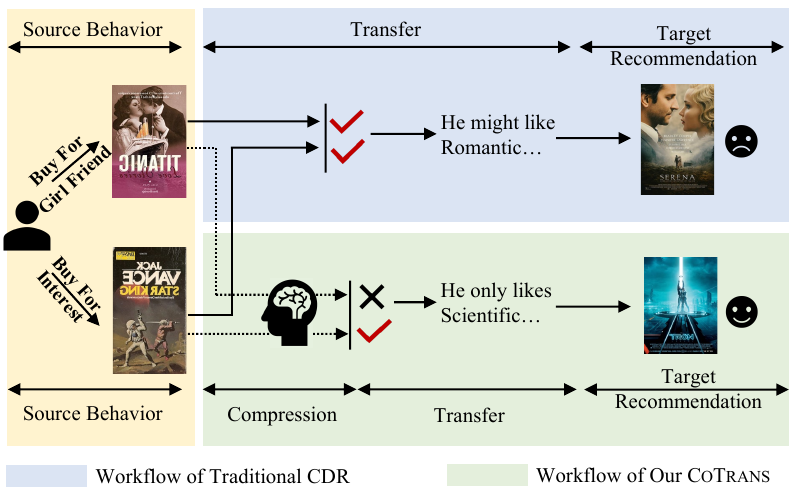}
  \caption{\textcolor{black}{Illustration of the workflow of our {\model} compared with tranditional CDR methods.}}
  \label{fig:intro-case}
\end{figure}

To address these headaches, researchers have recently resorted to the Cross-Domain Recommendation (CDR) that effectively transfers knowledge from informative scenarios (source domain) to improve the learning process of poor scenarios with sparse interactions (target domain).
Earlier work~\cite{zhao2023cross,liu2021leveraging,zhu2019dtcdr} aims at learning shared knowledge through embedding-level mapping and alignment across different domains, followed by a multi-task learning paradigm.
However, these methods commonly learn a unified mapping function shared by all users through a shallow architecture, thus resulting in inflexible representation transferring.
Further, more advanced architectures are proposed to establish correlations across different domains for more precise recommendations.
In particular, recent approaches towards CDR mainly focus on the following three aspects:
i) fine-grained sequential pattern transferring~\cite{li2022recguru,ma2022mixed} via a deep attention mechanism
ii)  effectively characterizing user-specific preference in CDR with meta-learning~\cite{zhu2021transfer,xu2022metacar} 
and iii) flexibly incorporating rich auxiliary information (\eg taxonomy) into CDR with graph learning~\cite{liu2020cross,zhao2019cross,guo2021gcn} and contrastive learning~\cite{yi2023contrastive,cao2022contrastive}.
Although substantial advancements have been made, most of them commonly overlook such an essential question ``\emph{For each user, does all knowledge from source domains deserve being transferred ?}"

Generally, most methods tailored for CDR assume that the entirety of information from the source domain is equally potential to be beneficial to the target domain. 
Unfortunately, it may not hold in most scenarios, especially for complex real-world applications, as interactions (\eg click or purchase) of a certain user towards items are not completely motivated by his/her interests. Taking Fig.~\ref{fig:intro-case} as an example, although a user has purchased two books, it is possible that only the science fiction genre truly aligns with his genuine interests.
This suggests that not all behavior history is reflective of users' actual preferences, even contaminated with extraneous noise stemming from accidental clicks or other non-intentional actions \cite{wang2021denoising,gao2022self}. Subsequently, blindly fitting the observational behavior data in the source domains without considering the inherent correlation with target domains may seriously deteriorate the effectiveness of CDR, as illustrated in the upper part of Fig.~\ref{fig:intro-case}.

To tackle this challenge, we aim to harness abundant behaviors from source domains to facilitate preference learning in target domains upon graph neural network architecture in a more principled way. That is, we anatomize the procedure of CDR into two subtasks: \emph{Compression} and \emph{Transfer}. 
Specifically, as shown in the bottom part of Fig.~\ref{fig:intro-case}, our workflow adaptively discard target-irrelevant behaviors from the source ``Book''domain (\ie \emph{Compression}) and preserve the core knowledge for the recommendation of the target ``Movie'' domain (\ie \emph{Transfer}).
We borrow the main idea of graph information bottleneck theory~\cite{wu2020graph,tishby2000information} (GIB) due to its ability to extract the minimum sufficient information --- which is closely aligned with the goal of CDR to maintain the efficient and relevant information flow from source to target domains. 
Although the graph information bottleneck holds great potential for enhancing CDR tasks, 
the solution is quite challenging. 
Firstly, the identification of important and relevant user behaviors in the source domain is inherently dependent on their behaviors within target domains, highlighting the necessity for the \emph{Compression} process to incorporate considerations of the target domain-specific contexts. 
Furthermore, in many scenarios, distinct domains may not have overlapping items~\cite{liu2020cross,han2023intra}, potentailly leading to completely different item feature space. Bridging this content gap is essential for enabling effective \emph{Transfer}.

To this end,  we propose an innovative knowledge enhanced cross-domain recommendation framework {\model} that effectively \underline{Co}mpresses important information in source domains and then \underline{Trans}fers to the target domains.
For the \textit{Compression} process, we creatively leverage the users' behaviors in the target domains to guide the graph information bottleneck, allowing the procedure to efficiently identify and prioritize information that is essential for the target domain.
As for the process of \textit{Transfer}, the purified source information, together with information from the target domain, contribute to the prediction task in both domains. Thus, more robust decisions could be made for the recommendation.
Besides, the content gap arising from non-overlapping items between distinct domains significantly hampers effective knowledge integration during both the \emph{Compression} and \emph{Transfer} processes. 
We have empirically validated this assertion (see Section~\ref{sec:ablation}) --- without an intermediary to facilitate knowledge propagation between domains, attempting to directly transfer knowledge from one domain to another could be ineffective or even harmful.
To eliminate the knowledge isolation between domains, we introduce an open-sourced Knowledge Graph~\cite{bordes2013translating} to infer the item embedding, serving as a bridge for effective knowledge propagation.

We summarize our main contributions as follows:
i) We reformulate the CDR task from the perspective of the Information Bottleneck and propose an innovative paradigm for CDR task, \ie \textit{Compression}  and \textit{Transfer}. To the best of our knowledge, we are the first effort to reveal the intrinsic of CDR task from the GIB perspective with full consideration of behaviors in target domains.
ii) We introduce a knowledge enhanced encoder, which employs a knowledge graph (KG) as the intermediary between different domains, thus overcoming the information heterogeneity between different domains that hinders knowledge propagation.
iii)  We compare {\model} to extensive algorithms including both single-domain and cross-domain recommenders, achieving significant improvements on all CDR datasets.

\section{Preliminary \label{sec:pre}}


\subsection{Problem Statement}


Cross-Domain Recommendation (CDR) is designed to improve the recommendation performance on the target domain via utilizing data from another source domain. With the user set $\mathcal{U}^{\mathcal{S}}$ ($\mathcal{U}^{\mathcal{T}}$) and the item set  $\mathcal{I}^{\mathcal{S}}$ ($\mathcal{I}^{\mathcal{T}}$) of the source (target) domain $\mathcal{S}$ ($\mathcal{T}$), we define the implicit feedback matrix as $\mathbf{Y}^{\mathcal{S}} \in \{0, 1\}^{|\mathcal{U}^{\mathcal{S}}| \times |\mathcal{I}^{\mathcal{S}}|}$ ($\mathbf{Y}^{\mathcal{T}} \in \{0, 1\}^{|\mathcal{U}^{\mathcal{T}}| \times |\mathcal{I}^{\mathcal{T}}|}$), where $y^{\mathcal{S}}_{u,i} = 1$ ($y^{\mathcal{T}}_{u,i} = 1$) if and only if there exists a interaction between the user $u$ and item $i$ in the source (target) domain. The goal of CDR is to learn the user preference in the source domain $\mathcal{S}$ and transfer it into the target domain $\mathcal{T}$ to improve the recommendation performance on the target domain.
In this paper, we study the situation that has been most widely studied~\cite{singh2008relational,zhao2019cross,liu2020cross,li2009can} where a set of users is shared in both domains $\mathcal{U}^{\mathcal{S}} = \mathcal{U}^{\mathcal{T}}$, while items are completely non-overlapping across domains (\ie $\mathcal{I}^{\mathcal{S}} \cap \mathcal{I}^{\mathcal{T}} = \varnothing$). 

Recently, Graph-based methods ~\cite{liu2019geniepath,yu2022graph} have been demonstrated to be highly effective in generating high-quality user and item representations from historical interactions, thereby exhibiting prominent performance in recommendation. Following previous works ~\cite{zhao2023cross,zhao2019cross}, this study also focuses on graph-based recommendation. We represent interaction data as a user-item bipartite graph, \ie $\mathcal{G}^{\mathcal{S}} = \{\mathcal{V}^{S}, \mathcal{E}^{S}\}$ ($\mathcal{G}^{\mathcal{T}} = \{\mathcal{V}^{\mathcal T}, \mathcal{E}^{\mathcal T}\}$) with the node set $\mathcal{V}^{\mathcal S} = \mathcal{U}^{\mathcal{S}} \cup  \mathcal{I}^{\mathcal{S}}$ ($\mathcal{V}^{\mathcal T} = \mathcal{U}^{\mathcal{T}} \cup  \mathcal{I}^{\mathcal{T}}$) and the edge set $\mathcal{E}^{S} = \{(u,i)| u \in \mathcal{U}^{\mathcal{S}}, i \in \mathcal{I}^{\mathcal{S}}, y^\mathcal{S}_{u,i} = 1\}$ ($\mathcal{E}^{T} = \{(u,i)| u \in \mathcal{U}^{\mathcal{T}}, i \in \mathcal{I}^{\mathcal{T}}, y^\mathcal{T}_{u,i} = 1\}$).

\subsection{Information Bottleneck}
Learned from information theory, the identification of relevant information to retain and irrelevant information to discard from input data holds significance. 
The information bottleneck (IB) theory~\cite{tishby2000information} offers a systematic and principled framework to address this challenge. 
By compressing the source random variable, IB aims to retain the information essential for accurately predicting the target random variable while disregarding any information that is not relevant to the target variable.

\begin{definition}{Informantion Bottleneck.}
Given the original data $\bm{X}$ with label $\mathbf{Y}$, IB is to obtain a compact and effective representation $\bm{Z}$ of $\bm{X}$.
The objective of the IB principle is as follows:
\begin{equation}
\label{eq:IB}
    \mathop{min}\limits_{\bm{Z}} -I(\mathbf{Y};\bm{Z})+\beta I(\bm{X};\bm{Z}), 
\end{equation}
where $I(\cdot; \cdot)$ denotes the mutual information of two random variables, $\beta$ is a Lagrangian multiplier for balancing the two terms. 
\end{definition}

Recently, the information bottleneck principle has also been employed to learn an essential and minimal graph structure.
\begin{definition}{Graph Information Bottleneck.}
    Given a graph $\mathcal{G} = \{\mathcal{V}, \mathcal{E}\}$ and its label information $\mathbf{Y}$, the graph information principle aims at discovering the optimal graph   $\mathcal{\hat{G}} = \{\mathcal{\hat{V}}, \mathcal{\hat{E}}\}$ through: 
\begin{equation}
\label{eq:GIB}
    \mathop{\min}\limits_{\mathcal{\hat{G}}}-I(\mathbf{Y};\mathcal{\hat{G}})+\beta I(\mathcal{G};\mathcal{\hat{G}}),
\end{equation}
where $\mathcal{\hat{V}}$ and $\mathcal{\hat{E}}$ represent the task-relevant set of nodes and edges of the original graph $\mathcal{G}$ respectively.
\end{definition}

\section{Methodology \label{sec:method}}


\begin{figure}
\centering
  \includegraphics[width=0.49\textwidth]{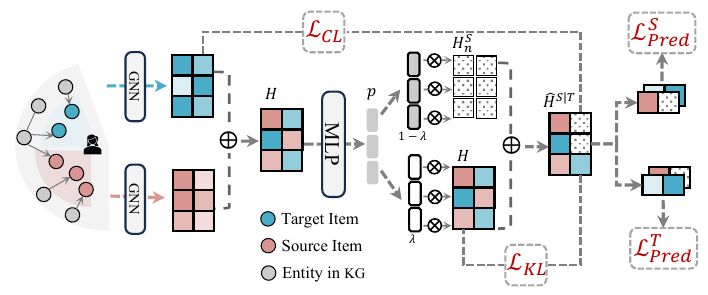}
  \caption{The overall architecture of the proposed {\model}.}
  \label{fig:framework}
\end{figure}


In this section, we present our proposed approach {\model}, an innovative cross-domain recommendation framework guided by the graph information bottleneck that \underline{Co}mpresses and then \underline{Trans}fers user preference from source domains to facilitate the recommendation performance in target domains.

\subsection{Cross-domain Recommendation under Graph Information Bottleneck \label{sec:method_1}}
We first reformulate the CDR task from the perspective of the graph information bottleneck and then demystify the underlying reasons.

\subsubsection{Formulation\label{sec:formulation}}
In our study, we are interested in learning the core behavior sub-graph of a user from the source domain $\mathcal{S}$ with the careful consideration of his/her interactions in the target domain $\mathcal{T}$, which efficiently encapsulates user preference within source domains, ensuring that only the most relevant and impactful knowledge is retained.
Consequently, we establish the fundamental bridge between the graph information bottleneck and CDR as follows:

\begin{definition}{Graph Information Bottleneck Towards CDR.}
Given a user's behavior graph $\mathcal G^{\mathcal{S}}$ and feedback matrix $\mathbf{Y}^{\mathcal{S}}$ in the source domain $\mathcal{S}$, paired with his/her behavior graph $\mathcal G^{\mathcal{T}}$ and $\mathbf{Y}^{\mathcal{T}}$ in target domain $\mathcal{T}$, our goal is to obtain most relevant/important behaviors towards target domains, denoted as $\mathcal{\hat{G}}^{\mathcal{S}|\mathcal{T}}$,
    \begin{equation}
\label{eq:opt_overall}
    \mathop{\min}_{\mathcal{\hat{G}}^{\mathcal{S}|\mathcal{T}}}\
    \underbrace{\beta I(\mathcal{G}^{\mathcal{S}};\mathcal{\hat{G}}^{\mathcal{S}|\mathcal{T}}|\mathcal{G}^{\mathcal{T}})}_{\textit{Compression}}\underbrace{-I(\mathbf{Y}^{\mathcal{T}};\mathcal{\hat{G}}^{\mathcal{S}|\mathcal{T}}|\mathcal{G}^{\mathcal{T}})
    -I({\mathbf{Y}}^{\mathcal{S}};\mathcal{\hat{G}}^{\mathcal{S}|\mathcal{T}}|\mathcal{G}^{\mathcal{T}})}_{\textit{Transfer}}.
\end{equation}
\end{definition}
Intuitively, in the settings of CDR, the graph information bottleneck aims to learn the core behavior of the source domains via the following \emph{Compression} and \emph{Transfer} procedures:
\begin{itemize}[leftmargin=*]
    \item We introduce the users' behaviors in the target domains into the information bottleneck for a more rational compression, allowing our framework to prioritize information that is essential for the target domain. In particular, guided by signals in the target domains, target-irrelevant information from the source domains is efficiently discarded by minimizing the term  $I(\mathcal{G}^{\mathcal{S}};\mathcal{\hat{G}}^{\mathcal{S}|\mathcal{T}}|\mathcal{G}^{\mathcal{T}})$.
    \item In the settings of CDR, signals pertinent to the recommendation task should be effectively maintained across both the source and target domains since the efficacy of the CDR system hinges on its ability to leverage informative cues from one domain to enhance its predictions in another. Therefore, the transfer module guarantees the information preservation in the source domain by maximizing the term $I(\mathbf{Y}^{\mathcal{S}};\mathcal{\hat{G}}^{\mathcal{S}|\mathcal{T}}|\mathcal{G}^{\mathcal{T}})$ and performance effective transfer of task-relevant signals to target domains by maximizing $I({\mathbf{Y}}^{\mathcal{T}};\mathcal{\hat{G}}^{\mathcal{S}|\mathcal{T}}|\mathcal{G}^{\mathcal{T}})$
\end{itemize}

\subsubsection{Why it works ?}
The innovative formulation allows us to examine the CDR task through the lens of information theory, potentially inducing a robust approach that is resilient against noise and superfluous structures within real-world recommender systems. In particular, we provide a  theoretical guarantee as follows:

\begin{lemma}
\label{lemma}
Given a pair of behavior graphs $(\mathcal{G}^\mathcal{S},\mathcal{G}^\mathcal{T})$ for both domains and corresponding feedback matrix $\{\mathbf{Y}^{\mathcal{S}},\mathbf{Y}^\mathcal{T}\}$, let $\mathcal{\tilde{G}}^\mathcal{S}$ represent users' redundant behaviors in the source domain that are irrelevant to their true preference.
Then the following inequality holds:
\begin{equation}
\label{eq:lemma}
    \begin{split}
        I(\mathcal{\hat{G}}^{\mathcal{S}|\mathcal{T}};\mathcal{\tilde{G}}^\mathcal{S}|\mathcal{G}^\mathcal{T}) & \leqslant 
        -I(\mathbf{Y}^{\mathcal{T}};\mathcal{\hat{G}}^{\mathcal{S}|\mathcal{T}}|\mathcal{G}^{\mathcal{T}})
        -I(\mathbf{Y}^{\mathcal{S}};\mathcal{\hat{G}}^{\mathcal{S}|\mathcal{T}}|\mathcal{G}^{\mathcal{T}}) \\
        & + I(\mathcal{G}^\mathcal{S};\mathcal{\hat{G}}^{\mathcal{S}|\mathcal{T}}|\mathcal{G}^{\mathcal{T}}).
    \end{split}
\end{equation}
\end{lemma}
In short, the lemma clearly tells that our reformulated objective of graph information bottleneck for CDR (\ie Eq.~\eqref{eq:opt_overall}) is an upper bound of conditional mutual information $I(\mathcal{\hat{G}}^{\mathcal{S}|\mathcal{T}};\mathcal{G}^{\mathcal{S}}_{n}|\mathcal{G}^\mathcal{T})$ when $\beta = 1$.
In other words, optimizing objective in Eq.~\eqref{eq:opt_overall} is equivalent to minimizing the mutual information between the core sub-graph $\mathcal{\hat{G}}^{\mathcal{S}|\mathcal{T}}$ and noisy structure $\mathcal{\tilde{G}}^\mathcal{S}$. Hence, it provides a theoretical guarantee that our formulation leads to the noise invariance property for CDR by discarding the information from the source domains that have no benefit to the final recommendation task in the target domain. \textcolor{black}{A detailed proof is given in Appendix~\ref{sec:lemma_appendix_1}.}

\subsection{Instantiation of {\model} \label{sec:method_2}}
To this end, the reformed {\model} under Graph Information Bottleneck has provided a novel and advanced solution for CDR: 
(1) {\model} operates a customized compression upon information in the source domain, enabled by discerning personal behaviors in the target domain.
(2) {\model} promises a stable knowledge transfer by preserving the task-relevant signal in both domains. 
As a whole, the engagement of compressed source information also vouches for a more controllable transfer procedure.
In this section, we will give a deeper insight into {\model} and detail how it works to operate the compression and transfer procedures.
Specifically, we first introduce the knowledge-enhanced encoder that eliminates the isolation between distinct domains.
Subsequently, we will instantiate the compression and transfer procedures of {\model}, and reveal the mechanism behind them.

\subsubsection{Knowledge-enhanced Encoder.}
As mentioned in Section~\ref{sec:Intro}, 
the information in the source domain cannot directly serve the target recommendation, since the items are not shared between the source and target domains, causing the distributions of aggregated representations from distinct domains not compatible.
To overcome this problem, we design an easily deployed encoder to produce generalized representations by implanting a knowledge graph (KG) with entity set $\mathcal{B}$ between distinct domains. 
Specifically, we define the correlation between the entities as a matrix $\mathbf{Y}_{e \rightarrow e}$, and provide the mapping between items and entities through the matrix $\mathbf{Y}_{i \rightarrow e}$. For the encoding towards the source domain $\mathcal{S}$, we organize them with the  user-item bipartite graph $\mathcal{G}^{\mathcal{S}}$ (denoted as $\mathbf{Y}^{\mathcal{S}}_{u \rightarrow i}$ below for brevity) as the adjacency matrix $\mathbf{A}_{\mathcal{G}}^\mathcal{S}$:
\begin{equation}
   \mathcal{A}_{\mathcal{G}}^\mathcal{S} =  \begin{bmatrix}
    \mathbf{0} & \mathbf{Y}^{\mathcal{S}}_{u \rightarrow i} & \mathbf{0}\\
    \mathbf{Y}^{\mathcal{S}}_{i \rightarrow u} & \mathbf{0} &  \mathbf{Y}_{i \rightarrow e}\\
    \mathbf{0} & \mathbf{Y}_{e \rightarrow i} & \mathbf{Y}_{e \rightarrow e} \\
    \end{bmatrix},
\end{equation}
where $\mathbf{Y}^{\mathcal{S}}_{i \rightarrow u}$ and $\mathbf{Y}_{e \rightarrow i}$ is the 
transpose of matrix $\mathbf{Y}^{\mathcal{S}}_{u \rightarrow i}$ and $\mathbf{Y}_{i \rightarrow e}$, respectively. 
In our study, we adopt the simple and effective convolution from  LightGCN~\cite{he2020lightgcn} to derive the generalized and informative representations, which utilizes weighted sum aggregators and omits feature transformation and nonlinear activation functions.
\begin{equation}
    \label{eq:gnn}
    \mathbf{E}^{\mathcal{S} (l)} = ({\mathbf{D}^\mathcal{S}_\mathcal{G}}^{-\frac{1}{2}} \mathbf{A}_{\mathcal{G}}^\mathcal{S} {\mathbf{D}^\mathcal{S}_\mathcal{G}}^{-\frac{1}{2}} ) \mathbf{E}^{\mathcal{S} (l - 1)},
\end{equation}
where $\mathbf{D}^\mathcal{S}_\mathcal{G}$ is a diagonal matrix, where each each entry ${d^\mathcal{S}_\mathcal{G}}_{i,i}$ denotes the number of nonzero entries in the $i$-th row of $\mathbf{A}_{\mathcal{G}}^\mathcal{S}$. And $\mathbf{E}^{\mathcal{S} (0)} = [\mathbf{E}_u^{\mathcal{S} (0)}, \mathbf{E}_i^{\mathcal{S} (0)}, \mathbf{E}_e^{(0)}]$ is the initial input of our encoder where $\mathbf{E}_u^{\mathcal{S} (0)}$ and $\mathbf{E}_e^{(0)}$ are initialized with learnable parameters, $\mathbf{E}_i^{\mathcal{S} (0)}$ is set as the zero matrix since item information cannot be shared across different domains.
At last, we extract the $L$-th layer output of users and items as the generalized representations of {\model}, denoted as $\mathbf{Z}^{\mathcal{S}} = [\mathbf{E}_u^{\mathcal{S} (L)}, \mathbf{E}_i^{\mathcal{S} (L)}]$. Analogously, the generalized representations towards target domain $\mathcal{T}$ can be deduced.


\subsubsection{Instantiation of Compression.}
To probe the most relevant user behavior towards the target domain, the first term of Eq.~\eqref{eq:opt_overall} is required to be minimized. 
In special, we decompose it into two terms based on the chain rule of mutual information~\cite{fawzi2015quantum} as follows:
\begin{equation}
\label{eq:opt_pred_3}
I(\mathcal{G}^{\mathcal{S}};\mathcal{\hat{G}}^{\mathcal{S}|\mathcal{T}}|\mathcal{G}^\mathcal{T})=I(\mathcal{\hat{G}}^{\mathcal{S}|\mathcal{T}};\mathcal{G}^\mathcal{S},\mathcal{G}^\mathcal{T})-I(\mathcal{\hat{G}}^{\mathcal{S}|\mathcal{T}};\mathcal{G}^\mathcal{T}),
\end{equation}
where minimizing $I(\mathcal{\hat{G}}^{\mathcal{S}|\mathcal{T}};\mathcal{G}^\mathcal{S},\mathcal{G}^\mathcal{T})$ supposes to integrate the knowledge $(\mathcal{G}^\mathcal{S},\mathcal{G}^\mathcal{T})$ from both the source and target domains and compress it into an optimal source subgraph $\mathcal{\hat{G}}^{\mathcal{S}|\mathcal{T}}$.
Meanwhile, maximizing $I(\mathcal{\hat{G}}^{\mathcal{S}|\mathcal{T}},\mathcal{G}^\mathcal{T})$ requires the compressed $\mathcal{\hat{G}}^{\mathcal{S}|\mathcal{T}}$ should keep as much the target-relevant information as possible.
A more thorough philosophy behind the above two terms is, that a rational compression upon source behaviors should be fulfilled with the perception of information from both the source and target domains.

\subsubsection{Instantiation of Transfer.}
To transfer the exquisite knowledge from the compressor into the target domain for superior recommendation, the rest two terms in Eq.~\eqref{eq:opt_overall} need to be minimized jointly. 
Following the chain rule of mutual information, the first term $-I( \mathbf{Y}^{\mathcal{T}};\mathcal{\hat{G}}^{\mathcal{S}|\mathcal{T}}|\mathcal{G}^\mathcal{T})$ can be decomposed as follows:
\begin{equation}
\label{eq:opt_pred_1}
    -I( \mathbf{Y}^{\mathcal{T}};\mathcal{\hat{G}}^{\mathcal{S}|\mathcal{T}}|\mathcal{G}^\mathcal{T})=-I( \mathbf{Y}^{\mathcal{T}};\mathcal{\hat{G}}^{\mathcal{S}|\mathcal{T}},\mathcal{G}^\mathcal{T})+I( \mathbf{Y}^{\mathcal{T}};\mathcal{G}^\mathcal{T}).
\end{equation}
Previous work~\cite{lee2023conditional} has proven that minimizing $I( \mathbf{Y}^{\mathcal{T}};\mathcal{G}^\mathcal{T})$ deteriorates the model performance on the final task. 
Thus we only consider optimize $-I( \mathbf{Y}^{\mathcal{T}};\mathcal{\hat{G}}^{\mathcal{S}|\mathcal{T}},\mathcal{G}^{\mathcal{T}})$ in this work.
Then we obtain the upper bound of $-I( \mathbf{Y}^{\mathcal{T}};\mathcal{\hat{G}}^{\mathcal{S}|\mathcal{T}},\mathcal{G}^\mathcal{T})$ from following proposition:
\begin{equation}
\label{eq:opt_pred_1_1}
    -I( \mathbf{Y}^{\mathcal{T}};\mathcal{\hat{G}}^{\mathcal{S}|\mathcal{T}},\mathcal{G}^\mathcal{T}) \leqslant \mathbb{E}_{\mathcal{\hat{G}}^{\mathcal{S}|\mathcal{T}},G^{\mathcal{T}}, \mathbf{Y}^{\mathcal{T}}}[-\log p_{\theta}( \mathbf{Y}^{\mathcal{T}}|\mathcal{\hat{G}}^{\mathcal{S}|\mathcal{T}},\mathcal{G}^\mathcal{T})],
\end{equation}
where 
$p_{\theta}$ is the variational approximation of $p( \mathbf{Y}^{\mathcal{T}}|\mathcal{\hat{G}}^{\mathcal{S}|\mathcal{T}},\mathcal{G}^\mathcal{T})$, which could be modeled as a predictor that takes both the compressed source subgraph and users' original behaviors in the target domain as input. 
As the core atom of the transfer procedure, this term induces the compressive information $\mathcal{\hat{G}}^{\mathcal{S}|\mathcal{T}}$, in tandem with the signal in the target domain, to promote the target recommendation.
Besides, as mentioned in Section~\ref{sec:formulation}, all signals that are pertinent to the recommendation task should be preserved to keep the CDR system at a superior state.
Thus feedback signals $\mathbf{Y}^{\mathcal{S}}$ from the source domain should also participate in the transfer procedure, as an additional supervision for the learning process of $\mathcal{\hat{G}}^{\mathcal{S}|\mathcal{T}}$.
Similar with the formulation in Eq.~\eqref{eq:opt_pred_1_1}, this term could also be instantiated
by minimizing its upper bounds with another predictor $q_{\phi}$.
The proof for Eq~\eqref{eq:opt_pred_1_1} will be detailed in Appendix~\ref{sec:lemma_appendix_2}.

Up to now, we have gained a profound see into the mechanism of {\model}.
Given an airscape upon the whole model, we observe that {\model} strives to perform a careful knowledge \textit{compression and transfer} for the CDR task. 
The perspective of the graph information bottleneck also supplies a decomposition to the CDR task and assists the recommender in making more sensible decisions. 

\subsection{Optimization of {\model} \label{sec:method_3}}
In this section, we detail the optimization for each part of {\model}.


\subsubsection{Optimization for Compression. \label{sec:opt_comp}}
To optimize the first term $I(\mathcal{\hat{G}}^{\mathcal{S}|\mathcal{T}};\mathcal{G}^\mathcal{S},\mathcal{G}^\mathcal{T})$ in Eq.~\eqref{eq:opt_pred_3}, inspired by the previous work~\cite{yu2022improving} that minimizes $I(\mathcal{G}^{IB};\mathcal{G})$ by injecting noise into node representations, 
we first introduce the merged representations $\mathbf{H}$ that contains the information of user behaviors from both source and target domains:
\begin{equation}
\label{eq:merged_H}
   \mathbf{H}= \mathbf{E}^\mathcal{S}+ \mathbf{E}^\mathcal{T},
\end{equation}
where $ \mathbf{E}^\mathcal{S}, \mathbf{E}^\mathcal{T}$ are generated using separate GNN models, aggregating users' high-order interaction in the source and target domains, respectively.
Based on the merged representations $\mathbf{H}$, we compress them into the most informative cache $\mathbf{\hat{H}}_i^{\mathcal{S}|\mathcal{T}}$ by replacing the devil information with random noise. 
More precisely, given a user $i$’s integrated embedding $\mathbf{H}_i$, we calculate a probability $p_i$ with MLP, \ie $p_i=MLP(\mathbf{H}_i)$.
With the calculated probability $p_i$, we contaminate the representation $\mathbf{H}_i$ of user $i$ with noise $\epsilon$:
\begin{equation}
\label{eq:opt_pred_3_noise}
\mathbf{\hat{H}}_i^{\mathcal{S}|\mathcal{T}} = \lambda_i \mathbf{H}_i+(1-\lambda_i)\epsilon, 
\end{equation}
where $\epsilon \sim N(\mu_{\mathbf{H}},\sigma^2_{\mathbf{H}})$, $\lambda_i \sim Bernoulli(Sigmoid(p_i))$.
Note that $\mu_{\mathbf{H}},\sigma^2_{\mathbf{H}}$ are mean and variance of $\mathbf{H}$.
Such operation induces the information compression of $(\mathcal{G}^\mathcal{S},\mathcal{G}^\mathcal{T})$ into $\mathcal{\hat{G}}^{\mathcal{S}|\mathcal{T}}$ by substituting users' thoughtless behaviors with noise in the source domain.
Besides, we adopt gumbel sigmoid~\cite{maddison2016concrete} to the discrete random variable $\lambda$ for differentiable sampling process:
\begin{equation}
\label{eq:opt_pred_3_noise_2}
\lambda_i=Sigmoid(1/tlog[p_i/(1-p_i)]+log[m/(1-m)]), 
\end{equation}
where $t$ is the temperature hyperparameter and $m \sim Uniform(0,1)$.

\begin{myremark}
The value of $\lambda_i$ estimates the reliability of behaviors for user $i$ in the source domain. 
A small $\lambda_i$ indicates that the majority of source behaviors are irrelevant even harmful for recommendation in the target domain, so most of the trivial information in the source domain could be discarded. 
By contrast, a large $\lambda_i$ inspires the {\model} to be more convinced with the source knowledge $\mathbf{H}_i$ and encourages the GNN backbone to aggregate better representations for transferring.
\end{myremark}

Finally, we minimize the upper bound of $I(\mathcal{\hat{G}}^{\mathcal{S}|\mathcal{T}};\mathcal{G}^\mathcal{S},\mathcal{G}^\mathcal{T})$ as:
\begin{equation}
\label{eq:opt_pred_3_1}
    \begin{split}
    \mathop{\min}\
    I(\mathcal{\hat{G}}^{\mathcal{S}|\mathcal{T}};\mathcal{G}^\mathcal{S},\mathcal{G}^\mathcal{T}) & \leq\mathbb{E}_{\mathcal{G}^\mathcal{S},\mathcal{G}^\mathcal{T}}[-\frac{1}{2}logM+\frac{1}{2B}M+\frac{1}{2B}Q^2]\\
    &:=\mathcal{L}_{KL}(\mathcal{\hat{G}}^{\mathcal{S}|\mathcal{T}};\mathcal{G}^\mathcal{S},\mathcal{G}^\mathcal{T}),
    \end{split}
\end{equation}
where $M=\Sigma_{j=1}^{B}(1-\lambda_j)^2$, $Q=\frac{\Sigma_{j=1}^{B}\lambda_j(\mathbf{H}_j-\mu_{\mathbf{H}})}{\sigma_{H}}$. 
$B$ denotes the bach size.
We give a detailed proof of this inequality in Appendix~\ref{sec:lemma_appendix_3}.

As for the second term $-I(\mathcal{\hat{G}}^{\mathcal{S}|\mathcal{T}};\mathcal{G}^\mathcal{T})$ in Eq.~\eqref{eq:opt_pred_3}, we construct a contrastive loss to measure the difference between the compressed subgraph $\mathcal{\hat{G}}^{\mathcal{S}|\mathcal{T}}$ with the original information of $\mathcal{G}^\mathcal{T}$, denoted as $\mathcal{L}_{CL}(\mathcal{\hat{G}}^{\mathcal{S}|\mathcal{T}};\mathcal{G}^\mathcal{T})$, then we implement the optimization as follows:
\begin{equation}
\label{eq:opt_pred_3_2}
        \mathcal{L}_{CL} = -\frac{1}{B}\Sigma_{i=1}^Blog \frac{exp(sim(\mathbf{E}_i^{\mathcal{T}},\mathbf{\hat{H}}_i^{\mathcal{S}|\mathcal{T}})/\tau)}{\Sigma_{j=1}^B exp(sim(\mathbf{E}_j^{\mathcal{T}},\hat{\mathbf{H}}_i^{\mathcal{S}|\mathcal{T}})/\tau)},
\end{equation}
where $\bm{B}$ and $\tau$ denote the batch size and the temperature hyperparameter, respectively.

\begin{myremark}
The compression process takes an implicit way to obtain the optimal representations of the sub-structure of source domains, without explicitly probing into what the structure looks like. It could work well as the representation learning has already preserved the property of the optimal sub-structure, meanwhile ensuring the high efficiency of {\model}.

\end{myremark}

\subsubsection{Optimization for Transfer.}
Now we give a detail of the optimization of the transfer.
As for the input of predictors $p_{\theta},q_{\phi}$ involved in this procedure, on the user side, we first fuse the compressed $\mathbf{\hat{H}}^{\mathcal{S}|\mathcal{T}}$ and users' representations $\bm{\mathbf{E}}^{\mathcal{T}}$ in the target domain.
Then for the items, we use the generalized representations in the corresponding domain and employ an inner production as the predictor with the fused user representations:
\begin{equation}
\label{eq:opt_predictor}
        \hat{\bm{ \mathbf{Y}^{\mathcal{S}}}} = <\mathbf{\hat{H}}_i^{\mathcal{S}|\mathcal{T}}\oplus \mathbf{E}^{\mathcal{T}}_{i},\mathbf{E}^{\mathcal{S}}_j>,
        \hat{\bm{ \mathbf{Y}^{\mathcal{T}}}} = <\mathbf{\hat{H}}_i^{\mathcal{S}|\mathcal{T}}\oplus \mathbf{E}^{\mathcal{T}}_{i},\mathbf{E}^{\mathcal{T}}_k>,
\end{equation}
where $\oplus$ denotes the element-wise addition.
$\mathbf{E}^{\mathcal{S}}_j,\mathbf{E}^{\mathcal{T}}_k$ are the items' representations generated by GNN in the respective domain.
With the output $\bm{\hat{\mathbf{Y}}^{\mathcal{S}}},\bm{\hat{\mathbf{Y}}^{\mathcal{T}}}$ of predictors $p_{\theta},q_{\phi}$, we can minimize the upper bound of $-I( \mathbf{Y}^{\mathcal{T}};\mathcal{\hat{G}}^{\mathcal{S}|\mathcal{T}}|\mathcal{G}^\mathcal{T})$ and $-I( \mathbf{Y}^{\mathcal{S}};\mathcal{\hat{G}}^{\mathcal{S}|\mathcal{T}}|\mathcal{G}^\mathcal{T})$ by minimizing the prediction loss $\mathcal{L}^{\mathcal{S}}_{pred},\mathcal{L}^{\mathcal{T}}_{pred}$, respectively, which can be modeled as both the point-wise (\eg Cross-Entropy Loss) and pair-wise loss (\eg BPR Loss).

\paratitle{Final Objective.} 
Finally, we obtain the overall objective for {\model} as follows:
\begin{equation}
\label{eq:loss_overall}
        \mathcal{L}_{total} = \mathcal{L}_{pred}^{\mathcal{T}}+\alpha_1\mathcal{L}_{pred}^{\mathcal{S}} +\alpha_2\mathcal{L}_{KL} + \alpha_3 \mathcal{L}_{CL},
\end{equation}
where $\alpha_1, \alpha_2$ and $\alpha_3$ are the hyperparameters for weighting the importance of prediction or compression components.

\section{Experiment \label{sec:exp}}

\begin{table*}
	\caption{Overall performance comparison on three datasets. The best result is bolded and the runner-up is underlined.}
	\label{tab:performance}
\centering
\renewcommand{\arraystretch}{0.97}
\setlength{\tabcolsep}{2.0mm}{
	\begin{tabular}{ccccccccccccc}
		\toprule
        \multirow{3}{*}{Model} & \multicolumn{6}{c}{Book$\rightarrow$Movie} & \multicolumn{6}{c}{Movie$\rightarrow$Book} \\
        \cmidrule{2-7}\cmidrule(l){8-13}
        {} & \multicolumn{2}{c}{NDCG} & \multicolumn{2}{c}{HIT} & \multicolumn{2}{c}{MRR}  & \multicolumn{2}{c}{NDCG} & \multicolumn{2}{c}{HIT} & \multicolumn{2}{c}{MRR} \\
        \cmidrule{2-3}\cmidrule(l){4-5}\cmidrule(l){6-7}\cmidrule(l){8-9}\cmidrule(l){10-11}\cmidrule(l){12-13}
        {} & {$@10$} & {$@100$} & {$@10$} & {$@100$} & {$@10$} & {$@100$} & {$@10$} & {$@100$} & {$@10$} & {$@100$} & {$@10$} & {$@100$} \\
        \midrule
        {NMF}               & {1.0898} & {3.3566}  & {2.3132} & {14.2972} & {0.7275} & {1.0761} & {0.4192} & {1.2096} & {0.8541} & {5.0089} & {0.2892} & {0.4124}\\
        {LightGCN}          & {2.2490} & {4.9154}  & {4.3061} & {18.2562} & {1.6259} & {2.0473} & {0.9586} & {2.3456} & {1.9395} & {9.1637} & {0.6651} & {0.8957}\\
        {CCDR}              & {2.7001} & {5.7566}  & {5.2402} & {21.1299} & {1.6654} & {2.1656} & {1.1201} & {2.5210} & {2.1797} & {9.5374} & {0.7990} & {1.0175}\\
        {TIGER}             & {2.3964} & {5.0264}  & {4.2384} & {18.9773} & {1.5627} & {2.0893} & {0.9831} & {2.4239} & {1.9624} & {9.3654} & {0.6571} & {0.8096}\\
        {DDTCDR}            & {2.4591} & {5.4375}  & {4.8043} & {20.6317} & {1.6739} & {2.1631} & {0.9845} & {2.0293} & {1.9039} & {9.2171} & {0.7171} & {0.9279}\\
        {DTCDR}             & {1.4254} & {3.6956}  & {2.6961} & {14.6767} & {1.0365} & {1.3860} & {0.4131} & {1.2754} & {0.9075} & {5.4804} & {0.2657} & {0.3960}\\
        {GA-DTCDR}          & {1.9068} & {1.2871}  & {4.0138} & {17.8999} & {1.2871} & {1.7032} & {0.7001} & {1.8616} & {1.4502} & {7.5534} & {0.4751} & {0.6578}\\
        {CDRIB}          & \underline{2.8909} & \underline{6.1967}  & \underline{5.3139} & \underline{22.2776} & \underline{2.1020} & \underline{2.6192} & \underline{1.2099} & \underline{2.7789} & \underline{2.4377} & \underline{10.2783} & \underline{0.8416} & \underline{1.0960}\\
        {{\model}}          & \textbf{2.9286} & \textbf{6.2388}  & \textbf{5.5516} & \textbf{22.8826} & \textbf{2.1373} & \textbf{2.6579} & \textbf{1.4019} & \textbf{2.8501} & \textbf{2.8648} & \textbf{10.4982}& \textbf{0.9599} & \textbf{1.1850}\\

        \midrule
        \multirow{1}{*}{} & \multicolumn{6}{c}{CDs$\rightarrow$Movie} & \multicolumn{6}{c}{Movie$\rightarrow$CDs} \\
        \cmidrule{2-7}\cmidrule(l){8-13}
        {NMF}               & {1.5619} & {3.9176}  & {3.2232} & {15.7511} & {1.0743} & {1.4291} & {0.4746} & {1.0532} & {0.9933} & {3.9732} & {0.3170} & {0.4123}\\
        {LightGCN}          & \underline{2.1118} & {4.7192}  & {4.1354} & {17.7580} & \underline{1.4982} & {1.9111} & {0.9651} & {1.9318} & {1.7839} & {6.9532} & {0.7156} & {0.8590}\\
        {CCDR}              & {1.9311} & {4.6406}  & {3.7908} & {18.0215} & {1.3688} & {1.7913} & \textbf{1.5041} & \underline{2.5122} & \underline{2.5542} & \underline{7.7438} & \underline{1.1847} & \underline{1.3499}\\
        {TIGER}             & {1.8776} & {3.4652}  & {3.3432} & {14.3670} & {1.1263} & {1.5189} & {1.2303} & {2.2447} & {2.1285} & {7.3991} & {0.9544} & {1.1186}\\
        {DDTCDR}            & {2.2156} & \underline{4.9699}  & \textbf{4.6422} & \underline{19.1567} & {1.4893} & {1.9147} & {1.1782} & {2.2154} & {1.9461} & {7.2370} & {0.9444} & {1.1184}\\
        {DTCDR}             & {1.5402} & {3.8263}  & {3.0813} & {15.1632} & {1.0741} & {1.4272} & {0.5936} & {1.6375} & {1.1833} & {6.6459} & {0.4152} & {0.5787}\\
        {GA-DTCDR}          & {1.5711} & {3.8629}  & {3.2435} & {15.3456} & {1.0691} & {1.4191} & {1.1924} & {2.1985} & {1.9663} & {7.2572} & {0.9520} & {1.1097}\\
        {CDRIB}          & {2.1061} & {4.9037}  & {4.2368} & {18.7918} & {1.4668} & \underline{1.9172} & {1.0188} & {1.8603} & {1.7839} & {6.2031} & {0.7839} & {0.9156}\\
        {{\model}}          & \textbf{2.2215} & \textbf{5.1454}  & \underline{4.3787} & \textbf{19.7040} & \textbf{1.5668} & \textbf{2.0327} & \underline{1.4505} & \textbf{2.5365} & \textbf{2.6556} & \textbf{8.2708} & \textbf{1.1865} & \textbf{1.3619}\\
        \midrule
        \multirow{1}{*}{} & \multicolumn{6}{c}{Book$\rightarrow$CDs} & \multicolumn{6}{c}{CDs$\rightarrow$Book} \\
        \cmidrule{2-7}\cmidrule(l){8-13}

        {NMF}               & {0.3168} & {0.7581}  & {0.6690} & {3.0407} & {0.2131} & {0.2771} & {0.3740} & {0.8767} & {0.6892} & {3.3651} & {0.2791} & {0.3558}\\
        {LightGCN}          & {0.4007} & {1.0234}  & {0.7095} & {3.8922} & {0.4513} & {0.5221} & {0.8976} & {2.0163} & {1.7434} & {7.5613} & {0.6419} & {0.8215}\\
        {CCDR}              & \textbf{1.4476} & \underline{2.3363}  & \underline{2.4529} & {7.1153} & \textbf{1.1403} & \textbf{1.2790} & {0.9058} & {2.0742} & {1.7636} & {7.8451} & {0.6465} & {0.8348}\\
        {TIGER}             & {0.8980} & {1.8796}  & {1.6015} & {6.7302} & {0.6826} & {0.8405} & {0.6978} & {1.7520} & {1.5391} & {7.1352} & {0.4521} & {0.6125}\\
        {DDTCDR}            & {1.2184} & {2.3216}  & {2.0069} & \underline{7.6627} & {0.9785} & {1.1624} & {0.8600} & {1.9923} & {1.6623} & {7.6019} & {0.6194} & {0.7953}\\
        {DTCDR}             & {0.3750} & {1.0437}  & {0.8007} & {4.3149} & {0.2485} & {0.3535} & {0.3434} & {1.0430} & {0.7384} & {4.4573} & {0.2259} & {0.3318}\\
        {GA-DTCDR}          & {0.9104} & {1.5327}  & {1.5609} & {4.8449} & {0.7141} & {0.8101} & {0.5225} & {1.3592} & {0.9528} & {5.3720} & {0.3924} & {0.5211}\\
        {CDRIB}          & {1.0258} & {2.0623}  & {1.7636} & {7.0748} & {0.8001} & {0.9716} & \underline{1.0765} & \underline{2.2584} & \underline{1.8853} & \underline{7.9870} & \underline{0.8356} & \underline{1.0293}\\
        {{\model}}          & \underline{1.3952} & \textbf{2.5464}  & \textbf{2.4731} & \textbf{8.5141} & \underline{1.0634} & \underline{1.2439} & \textbf{1.3313} & \textbf{2.7020} & \textbf{2.4123} & \textbf{9.5682} & \textbf{1.0051} & \textbf{1.2216}\\
		\bottomrule
	\end{tabular}}
\end{table*}

\subsection{Experimental Settings}
\subsubsection{Datasets and Evaluation Metrics}


In this section, we construct three pairs of cross-domain recommendation datasets including Amazon Movies (AM) \& Amazon Books (AB), Amazon Movies (AM) \& Amazon CDs (AC), and Amazon Books (AB) \& Amazon CDs (AC).
For each pair of CDR datasets, the users are fully overlapped and we conduct the CDR task in both directions for all the methods. 
A detailed description of the datasets is given in Table ~\ref{tab:dataset}.
Considering the knowledge graph, we employ Freebase~\footnotemark, an online knowledge base containing massively structured triples, and extract a sub-graph from it with the entities linked by items that exist in all the CDR datasets as seeds to reduce computing costs.
Specifically, the sub-KG contains 3,599,000 entities with 32,796,605 edges and is planted into each pair of CDR datasets utilized in this paper.
\footnotetext{https://developers.google.com/freebase.}


\begin{table}
\renewcommand{\arraystretch}{0.8}
	\caption{The statistics of datasets.}
	\label{tab:dataset}
\centering
\setlength{\tabcolsep}{2.5mm}{
	\begin{tabular}{ccccc}
		\toprule
                \multicolumn{1}{c}{Datasets} & {Domain} & {Users} & {Items} & {Train} \\
                \midrule
                \multirow{2}{*}{AM \& AB} & {Movie}  & \multirow{2}{*}{11,240} & {16,100} & {119,915}  \\
                {} & {Book} & {} & {47,377} & {179,743}\\
                \midrule
                \multirow{2}{*}{AB \& AC} & {Book}  & \multirow{2}{*}{4,933} & {34,765} & {98,763}  \\
                {} & {CDs} & {} & {52,625} & {139,893}\\
                \midrule
                \multirow{2}{*}{AC \& AM} & {CDs}  & \multirow{2}{*}{4,933} & {52,625} & {139,893} \\
                {} & {Movie} & {} & {13,708} & {80,090}\\
            \bottomrule
        \end{tabular}}
\end{table}

To evaluate the recommendation performance of our {\model} and other baseline approaches, we follow previous works~\cite{zhuo2022tiger} and adopt the ranking-based evaluation strategy, \ie \textit{leave-one-out evaluation}.
Specifically, for each user, we randomly select two historical interactions from the target dataset, one for
validation, and the other for testing. The rest of the dataset is regarded as the training set.
The validation and test sets are shared for all models to ensure fair evaluations.
For each pair of datasets, we only keep users with more than three historical interactions in both domains.

\subsubsection{Competitors}
To verify the effectiveness of {\model}, we compare with both single- and cross-domain recommendation models.
We first select NMF~\cite{he2017neural} and LightGCN~\cite{he2020lightgcn}, two classical approaches for single-domain recommendations as the most straight baselines.
As for the cross-domain recommenders, we compare against TIGER~\cite{zhuo2022tiger}, DTCDR~\cite{zhu2019dtcdr}, GA-DTCDR~\cite{zhu2020graphical}, CCDR~\cite{xie2022contrastive} and DDTCDR~\cite{li2020ddtcdr}.
Note that TIGER is a CDR approach tailored for the cold-start scenario and we implement it on an oracle setting, \ie the model is trained on the training set of the target domain.
Besides, we additionally compare with another model that also incorporates an information bottleneck in the CDR task, called CDRIB~\cite{cao2022cross}.

\subsubsection{Implement Details.} 

%
Our framework is implemented using Pytorch. 
The embedding size for all representations is 32. 
We take Adagrad~\cite{duchi2011adaptive} with the initial learning rate 1e-3 for model optimization. 
For all the methods, we set the batch size to 4096 and the training maximum epoch to 100.
As for the set of hyper-parameters $\bm{\alpha} = \{\alpha_1,\alpha_2,\alpha_3\}$ involved in Eq~\eqref{eq:loss_overall}, 
we employ a grid search with an early stop strategy on NDCG@100 of the validation set to determine the best configuration. 
Table~\ref{tab:hyper-para} gives the settings of $(\alpha_1,\alpha_2,\alpha_3)$ on each pair of CDR datsets.
To ensure a fair comparison, all GNN-engaged CDR models apply the LightGCN as the backbone with the same structure, such as the number of layers.
To match an entity for each item, we feed the title of the item into the API released by Google~\footnote{https://developers.google.com/knowledge-graph.}, then the API returns the top-K relevant entities and we choose the first one as the corresponding entity.
Note that the extra computation cost introduced by Cotrans is $O(2|V|d^2)$, which is linear to the number of nodes, making it possible for efficient inference for a real-world recommender system.
\begin{table}
	\caption{Hyper-parameters settings for $\alpha_1,\alpha_2,\alpha_3$.}
	\label{tab:hyper-para}
\centering
\renewcommand{\arraystretch}{0.8}
\setlength{\tabcolsep}{0.4mm}{
	\begin{tabular}{c|c|c|c|c|c|c}
		\toprule
                \multicolumn{1}{c}{} & {AM$\rightarrow$AB} & {AB$\rightarrow$AM} & {AM$\rightarrow$AC} & {AC$\rightarrow$AM} & {AB$\rightarrow$AC} & {AC$\rightarrow$AB}\\
                \midrule
                \multicolumn{1}{c|}{$\alpha_1$} & {0.01} & {0.01} & {0.001} & {0.01} & {0.1} & {0.1} \\
                \midrule
                \multicolumn{1}{c|}{$\alpha_1$} & {1.0} & {0.01} & {1.0} & {1.0} & {1.0} & {1.0} \\
                \midrule
                \multicolumn{1}{c|}{$\alpha_1$} & {1.0} & {0.1} & {5.0} & {5.0} & {1.0} & {1.0} \\
            \bottomrule
        \end{tabular}}
\end{table}

\subsection{Performance Comparison \label{sec:exp_perfrom}}
Table~\ref{tab:performance} reports the performance of the three CDR datasets. 
Note that we conduct the CDR task on each dataset in a bidirectional setting, \ie the role of each sub-dataset would be switched between the target domain and source domain.
We have several observations: 

\paratitle{Comparison with Single-Domain Models:} 
(1) For single-domain methods, we observe that the GNN-based methods LightGCN consistently outperform NMF, which validates that the high-order information of collaborative filtering plays a key role in recommendations.
(2) {\model} significantly and consistently outperforms LightGCN by a considerable margin across all the metrics of all the datasets.
This indicates that the single-domain methods cannot make the optimal decision on the recommendation in the target domain without leveraging useful knowledge from other domains.

\paratitle{Comparison with Cross-Domain Models:} 
(1) Among the cross-domain methods, {\model} achieves the best performance in most cases. 
Such an encouraging performance gain indicates that {\model} could make a more elegant transfer of the knowledge from the source domain into the target recommendation task,
which is achieved by both the adoption of the general KG knowledge for gap mitigation between different domains and the adaptive source behavior compression conditioned on target knowledge.

\paratitle{CDR is Not Always Worked:} 
We surprisingly find that sometimes LightGCN could outperform CDR approaches by a large margin,
which demonstrates that an inappropriate transfer of source knowledge might cause a negative influence on recommendation tasks in the target domain.
This emphasizes a delicate behavior compression and transfer mechanism designed for the CDR task is necessary.

\subsection{Ablation Study  \label{sec:ablation}}
\begin{figure}
\centering
  \includegraphics[width=0.39\textwidth]{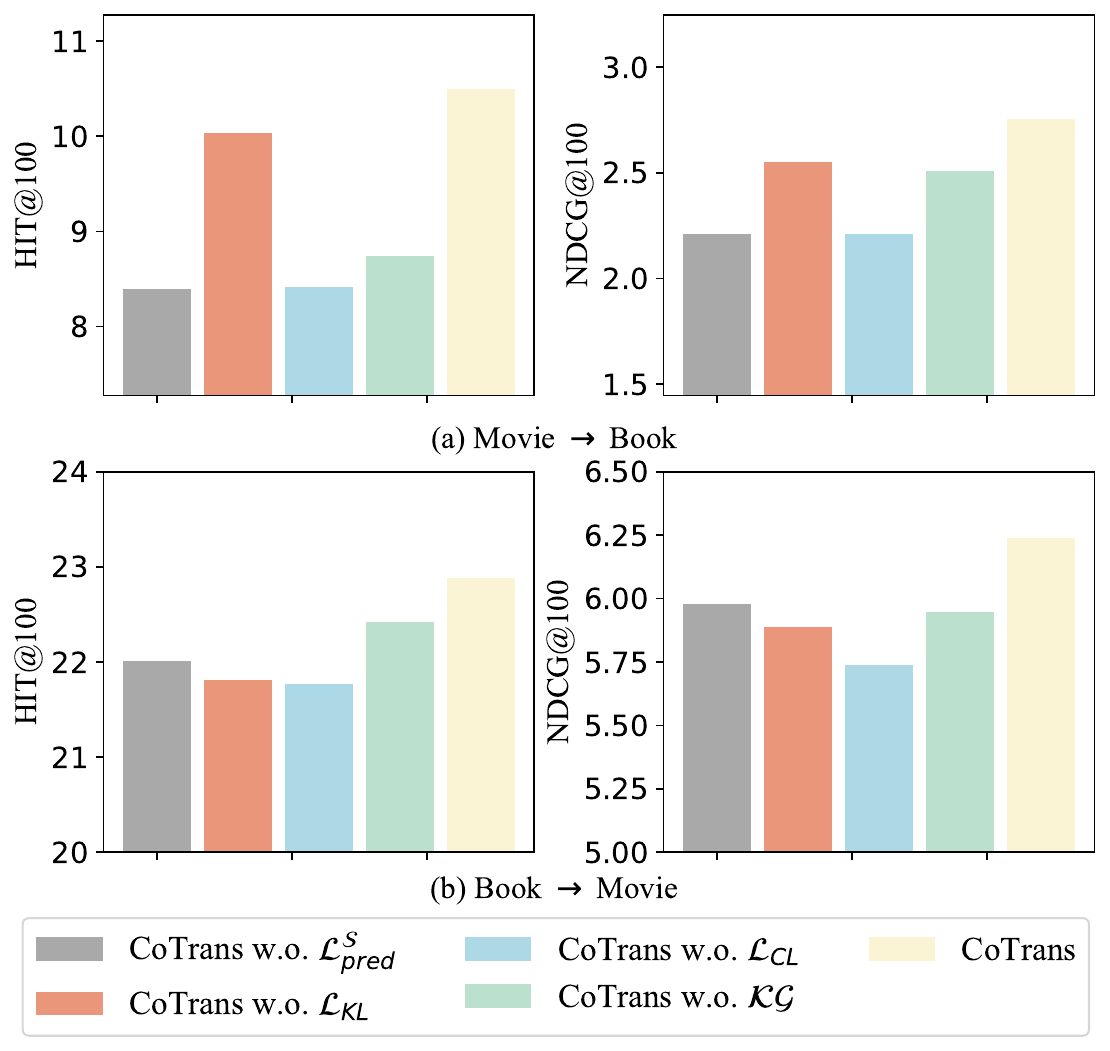}
  \caption{The results of ablation Study on the AM\&AB dataset.}
  \label{fig:ablation}
\end{figure}

To verify the benefit of the conditional compression module of our model, we first prepare four variants of {\model}:
i) {\model} without the objective $\mathcal{L}_{pred}^\mathcal{S}$, which is blind to the label information from the source domain.
ii) {\model} without the objective $\mathcal{L}_{KL}$, which loses the ability of compression for user behaviors in the source domain.
iii) {\model} without the objective $\mathcal{L}_{CL}$, which ignores the relevance between the compressed user behaviors in the source domain and his portrait in the target domain.
iiii) Besides, to estimate the necessity of KG incorporation for breaking the item isolation mentioned in Secion~\ref{sec:Intro}, we additionally run a full version of {\model} with simple ID-embedding initialization on the item side, which is completely irrelevant between distinct domains.
Due to the space limitation, we only report the experimental results on AM\&AB dataset in both directions.  


The results in Figure~\ref{fig:ablation} show that ignoring any module is not ideal, from which we have the following conclusions:
i) Label information from both domains is necessary for the final recommendation task.
ii) Knowledge compression for user behaviors in the source domain can promise the most relevant information to be transferred into the target domain.
iii) It is also essential to retain the reliable portrait of users derived from their behavior in the target domain.
iiii) Removing the intermediary that eliminates the isolation between separate domains would significantly harm the model performance.
In a word, each objective could assist the models in conducting more elaborate knowledge compression and transfer, subsequently enhancing the performance on downstream tasks.
Similar results are observed in other datasets.

\subsection{Sensitive Analysis}
\begin{figure}
\centering
  \includegraphics[width=0.42\textwidth]{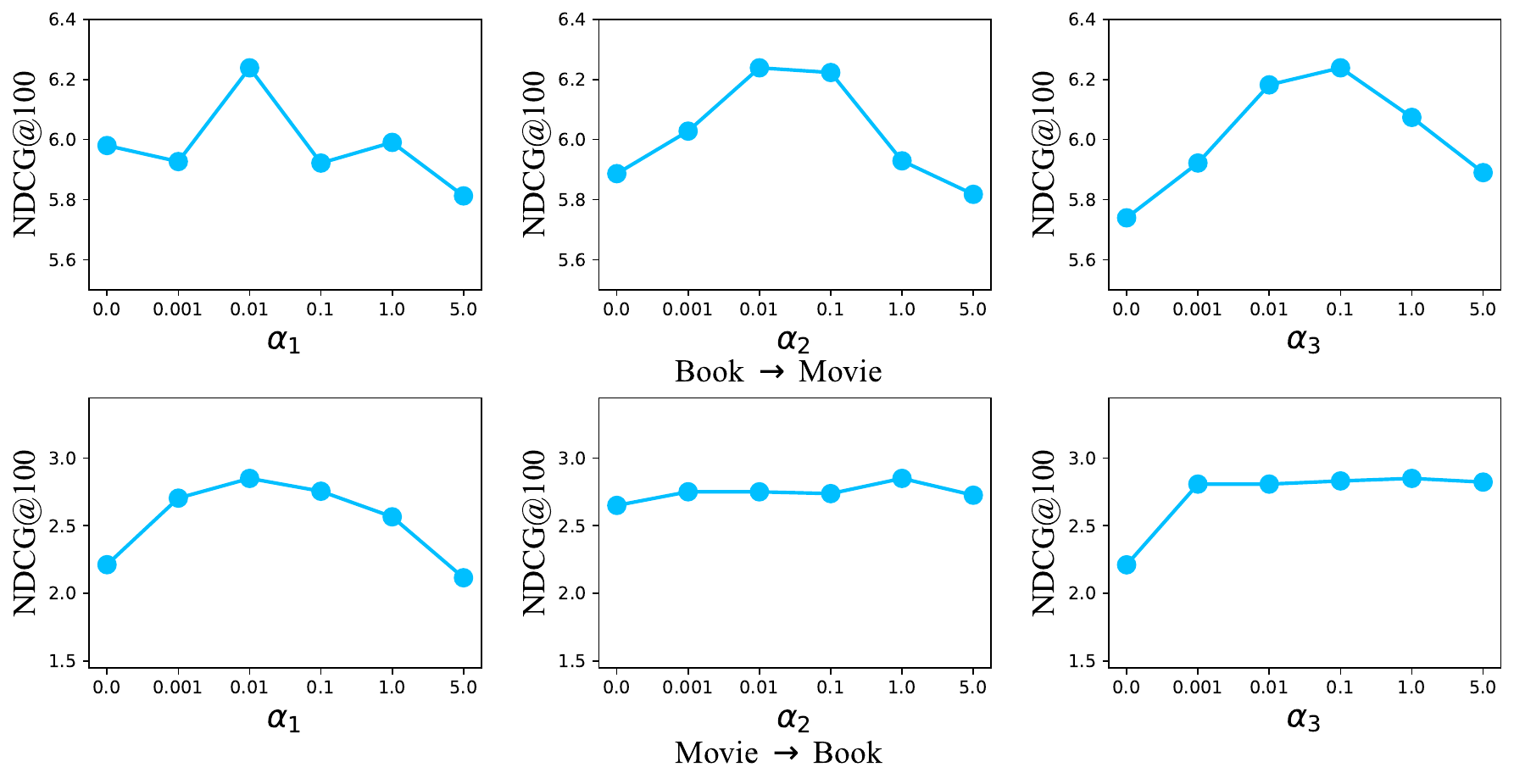}
  \caption{Sensitive Analysis on the AM\&AB dataset.}
  \label{fig:sensitive}
\end{figure}

As mentioned above, {\model} involves several important hyper-parameters, \ie balance factors $\alpha_1,\alpha_2$ and $\alpha_3$ in Eq.~\eqref{eq:loss_overall}.
Here, we investigate the impact of these parameters on the grid search strategy.
Due to space limitations, we just report the results on AM\&AB dataset in both directions.

From Fig.~\ref{fig:sensitive} we find: 
i) {\model} usually achieves the worst performance when any parameter of $\{\alpha_1,\alpha_2$,$\alpha_3\}$ is set to zero, demonstrating the indispensability of each module of {\model}.
ii) Large $\alpha_1$ causes apparent performance degradation, indicating that overreliance on the knowledge from the source domain is not encouraged and a trade-off between knowledge from different domains should be made.
Similar results are observed in other datasets.

\subsection{Robustness Analysis}


\begin{figure}[tbp]
\centering
  \includegraphics[width=0.42\textwidth]{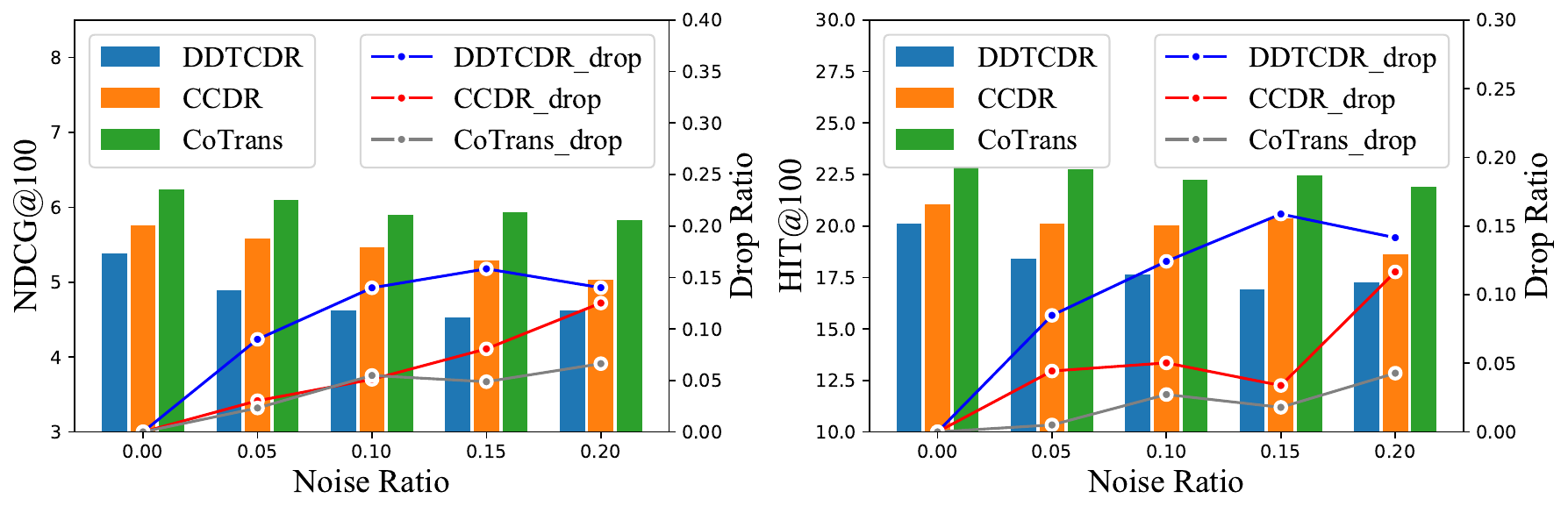}
  \caption{Performance with varying noise ratios.}
  \label{fig:robust}
\end{figure}

To verify {\model}'s robustness to the redundant interaction noises that existed in the source domain, we generate different proportions of negative interactions (\ie 5\%, 10\%, 15\%, 20\%) to contaminate the behavior graph in the source domain, and report the performance on the test set.
Besides, we chose CCDR and DDTCDR, two baselines that showed competitive performance in Section~\ref{sec:exp_perfrom} into the analysis.
Figure~\ref{fig:robust} shows the results on Amazon-Book (Source) and Amazon-Movie (Target), and the performance degradation ratio of the corresponding contaminated ratio.
It’s obvious that the more noise added to the source behavior, the worse performance all the models yield.
However, the performance degradation of {\model} is smoother and smaller than other approaches.
Specifically, from the left sub-figure of Figure~\ref{fig:robust}, when the noise ratio is small (from 0 to 0.1), the performance degradation of {\model} and CCDR is almost the same,
however, when further enlarging the ratio, the gap between {\model} and CCDR grows larger.
These observations imply that {\model} is robust to the noise that exists in the users' source behaviors, making it possible to concentrate on the most relevant source behaviors with the target recommendation task.


\section{Related Work \label{sec:rela}}
\subsection{Cross-Domain Recommendation}

CDR is supposed to solve the data sparsity problem that has confused recommender systems a lot by transferring knowledge across separate domains. 
Earlier work mainly solved this issue based on MF and its extension~\cite{hu2018conet,zhu2022personalized,lian2017cccfnet,hu2013personalized}.
In recent years, many more techniques have been employed in the CDR task including but not limited to meta-learning, graph learning, contrastive learning~\cite{zhu2020graphical,xie2022contrastive,tian2024privacy}, etc.
To name a few, GA-DTCDR~\cite{zhu2020graphical} first constructs two separate heterogeneous graphs from two domains and employs an element-wise attention mechanism to combine the embeddings of common users learned from both domains, thus improving the recommendation accuracy on each domain.

Despite their success, most CDR methods hold that the entirety of information from the source domain is equally beneficial to the target recommendation task, and should be preserved and transferred.
However, as we discussed in Section~\ref{sec:Intro}, it is inevitable for a user to interact with items that are not completely relevant to his/her internet interests.
It would be evil to the target recommendation if such inadvertent and random behaviors are transferred into the target domain.

\subsection{Graph Information Bottleneck}
Recently, many works~\cite{wu2020graph,yu2020graph,yu2022improving,wei2022contrastive} have proposed to extend the basic Information Bottleneck into graph data and generate a new branch of information theory named Graph Information Bottleneck (GIB), which has been widely applied to quite a few tasks.
For instance, GIB~\cite{yu2020graph} utilizes IB to regularize the structural information in the node classification task.
It employs the Shannon mutual information to measure how compressed and informative the subgraph distribution is. 
CDRIB~\cite{cao2022cross} is the only attempt that applies a vanilla information bottleneck (IB) approach to CDR.
However, this method just heuristically imposes the IB principle on the CDR scenario and the causality between the CDR task and IB theory is still a mystery.
Besides, CDRIB ignores users' target behaviors when optimizing the term for information minimality, 
whereas our \model~ defines a graph information bottleneck (CIB) for the CDR task, which incorporates target domain information as a condition when evaluating relevance. 
This target domain information is crucial in CDR, as it guides the extraction of relevant information from the source domain for transfer.

In summary, to the best of our knowledge, we are the first work that reveals the intrinsic mechanism of CDR tasks under a Graph Information Bottleneck perspective.
Moreover, the concretized procedures of \textit{Compression} and \textit{Transfer} make the CDR task easier to operate and be reformed for future study on this topic.

\section{Conclusion\label{sec:conc}}
In this paper, we have proposed the framework {\model} for the CDR task, which represents an attempt to \underline{Co}mpresse valuable sub-structure in source domains and then \underline{Trans}fers to the target domain.
In particular, following the theory of Graph Information Bottleneck, we operate the compression on the source behaviors with the perception of information from the target domain. 
Then to preserve all the task-relevant information during the procedure of transfer, we utilize the feedback signals from both domains for a more robust transfer of knowledge.
Besides, we employ a knowledge-enhanced encoder to narrow the gap between distinct domains and facilitate knowledge integration during both procedures of compression and transfer.
The extensive experiments conducted on three public datasets verify the effectiveness of {\model}.

\paratitle{Limitation}
As demonstrated in Section~\ref{sec:opt_comp}, current {\model} could compress the source behaviors by estimating the reliability of users' integrated representations.
Overall, the compression in the {\model} pipeline is built on the representation level (\ie directly embeds the optimal sub-structure into an embedding), which may lack of intuitive interpretability.
Obviously, more sophisticated compression might be achieved such as directly operating upon the structure of behavior graph $\mathcal{G}^{\mathcal{S}}$, which we leave as a future work.

\appendix
\section{APPENDIX}
\subsection{Proof of Lemma~\ref{lemma} \label{sec:lemma_appendix_1}}
Assuming that $\mathcal{G}^\mathcal{S},\mathcal{\hat{G}}^{\mathcal{S}|\mathcal{T}},\mathcal{\tilde{G}}^\mathcal{S},\mathcal{G}^\mathcal{T}$ and $\mathbf{Y}=(\mathbf{Y}^{\mathcal{S}},\mathbf{Y}^{\mathcal{T}})$ satisfy the markov condition:
\begin{equation}
    \begin{split}
    \label{eq:lemma_1}
        (\mathbf{Y},\mathcal{\tilde{G}}^\mathcal{S},\mathcal{G}^\mathcal{T})\rightarrow \mathcal{G}^\mathcal{S}\rightarrow\mathcal{\hat{G}}^{\mathcal{S}|\mathcal{T}}.
    \end{split}
\end{equation}
Following the data processing inequality, we have the following inequality:
\begin{equation}
    \begin{split}
    \label{eq:lemma_2}
        I(\mathcal{G}^\mathcal{S}&;\mathcal{\hat{G}}^{\mathcal{S}|\mathcal{T}}|\mathcal{G}^\mathcal{T}) = I(\mathcal{\hat{G}}^{\mathcal{S}|\mathcal{T}};\mathcal{G}^\mathcal{S},\mathcal{G}^\mathcal{T}) - I(\mathcal{\hat{G}}^{\mathcal{S}|\mathcal{T}};\mathcal{G}^\mathcal{T}) \\
            &\geq I(\mathcal{\hat{G}}^{\mathcal{S}|\mathcal{T}};\mathbf{Y},\mathcal{\tilde{G}}^\mathcal{S},\mathcal{G}^\mathcal{T}) - I(\mathcal{\hat{G}}^{\mathcal{S}|\mathcal{T}};\mathcal{G}^\mathcal{T}) \\
            &= I(\mathcal{\hat{G}}^{\mathcal{S}|\mathcal{T}};\mathcal{\tilde{G}}^\mathcal{S},\mathcal{G}^\mathcal{T}) + I(\mathcal{\hat{G}}^{\mathcal{S}|\mathcal{T}};\mathbf{Y}|\mathcal{\tilde{G}}^\mathcal{S},\mathcal{G}^\mathcal{T}) - I(\mathcal{\hat{G}}^{\mathcal{S}|\mathcal{T}};\mathcal{G}^\mathcal{T}) \\
            &= I(\mathcal{\hat{G}}^{\mathcal{S}|\mathcal{T}};\mathcal{\tilde{G}}^\mathcal{S}|\mathcal{G}^\mathcal{T}) + I(\mathcal{\hat{G}}^{\mathcal{S}|\mathcal{T}};\mathbf{Y}|\mathcal{\tilde{G}}^\mathcal{S},\mathcal{G}^\mathcal{T}).
    \end{split}
\end{equation}
Suppose that $\mathcal{\tilde{G}}^\mathcal{S}$ and $\mathbf{Y}$, $\mathcal{\tilde{G}}^\mathcal{S}$ and $\mathcal{G}^\mathcal{T}$, the joint $(\mathcal{\tilde{G}}^\mathcal{S},\mathcal{G}^\mathcal{T})$ and $\mathbf{Y}$ are independent, respectively. 
Then, for $I(\mathcal{\hat{G}}^{\mathcal{S}|\mathcal{T}};\mathbf{Y}|\mathcal{\tilde{G}}^\mathcal{S},\mathcal{G}^\mathcal{T})$, we have:
\begin{equation}
    \begin{split}
    \label{eq:lemma_3}
    I(\mathcal{\hat{G}}^{\mathcal{S}|\mathcal{T}};\mathbf{Y}|\mathcal{\tilde{G}}^\mathcal{S},\mathcal{G}^\mathcal{T}) &= H(\mathbf{Y}|\mathcal{\tilde{G}}^\mathcal{S},\mathcal{G}^\mathcal{T}) - H(\mathbf{Y}|\mathcal{\tilde{G}}^\mathcal{S},\mathcal{\hat{G}}^{\mathcal{S}|\mathcal{T}},\mathcal{G}^\mathcal{T}) \\
    &\geq H(\mathbf{Y}|\mathcal{G}^\mathcal{T}) - H(\mathbf{Y}|\mathcal{\hat{G}}^{\mathcal{S}|\mathcal{T}},\mathcal{G}^\mathcal{T}) \\
    &= I(\mathbf{Y};\mathcal{\hat{G}}^{\mathcal{S}|\mathcal{T}}|\mathcal{G}^\mathcal{T}) \\
    &= I(\mathbf{Y}^{\mathcal{S}},\mathbf{Y}^{\mathcal{T}};\mathcal{\hat{G}}^{\mathcal{S}|\mathcal{T}}|\mathcal{G}^\mathcal{T}) \\
    &= I(\mathcal{\hat{G}}^{\mathcal{S}|\mathcal{T}},\mathcal{G}^\mathcal{T};\mathbf{Y}^{\mathcal{T}},\mathbf{Y}^{\mathcal{S}}) - I(\mathcal{G}^\mathcal{T};\mathbf{Y}^{\mathcal{T}}, \mathbf{Y}^{\mathcal{S}}),
    \end{split}
\end{equation}
where the first term $I(\mathcal{\hat{G}}^{\mathcal{S}|\mathcal{T}},\mathcal{G}^\mathcal{T};\mathbf{Y}^{\mathcal{T}},\mathbf{Y}^{\mathcal{S}})$ can be formulated as: 
\begin{equation}
    \begin{split}
    \label{eq:lemma_4}
        I(\mathcal{\hat{G}}^{\mathcal{S}|\mathcal{T}},\mathcal{G}^\mathcal{T};\mathbf{Y}^{\mathcal{T}},\mathbf{Y}^{\mathcal{S}}) &= I(\mathcal{\hat{G}}^{\mathcal{S}|\mathcal{T}},\mathcal{G}^\mathcal{T};\mathbf{Y}^{\mathcal{T}}) + I(\mathcal{\hat{G}}^{\mathcal{S}|\mathcal{T}},\mathcal{G}^\mathcal{T};\mathbf{Y}^{\mathcal{S}}|\mathbf{Y}^{\mathcal{T}}) \\ 
        &= I(\mathcal{\hat{G}}^{\mathcal{S}|\mathcal{T}},\mathcal{G}^\mathcal{T};\mathbf{Y}^{\mathcal{T}}) + I(\mathcal{\hat{G}}^{\mathcal{S}|\mathcal{T}},\mathcal{G}^\mathcal{T};\mathbf{Y}^{\mathcal{S}}).
    \end{split}
\end{equation}
The second term $-I(\mathcal{G}^\mathcal{T};\mathbf{Y}^{\mathcal{T}}, \mathbf{Y}^{\mathcal{S}})$ can be decomposed as follows:
\begin{equation}
    \begin{split}
    \label{eq:lemma_5}
    -I(\mathcal{G}^\mathcal{T};\mathbf{Y}^{\mathcal{T}}, \mathbf{Y}^{\mathcal{S}}) &= -I(\mathcal{G}^\mathcal{T};\mathbf{Y}^{\mathcal{T}}) - I(\mathcal{G}^\mathcal{T};\mathbf{Y}^{\mathcal{S}}|\mathbf{Y}^{\mathcal{T}}) \\ 
    &=I(\mathcal{G}^\mathcal{T};\mathbf{Y}^{\mathcal{T}}) - I(\mathcal{G}^\mathcal{T};\mathbf{Y}^{\mathcal{S}})
    \end{split}
\end{equation}
By plugging Eq~\eqref{eq:lemma_4} and Eq~\eqref{eq:lemma_5} into Eq~\eqref{eq:lemma_3}, we have:
\begin{equation}
    \begin{split}
    \label{eq:lemma_6}
    &I(\mathcal{\hat{G}}^{\mathcal{S}|\mathcal{T}},\mathcal{G}^\mathcal{T};\mathbf{Y}^{\mathcal{T}},\mathbf{Y}^{\mathcal{S}}) - I(\mathcal{G}^\mathcal{T};\mathbf{Y}^{\mathcal{T}}, \mathbf{Y}^{\mathcal{S}}) \\ 
    &= I(\mathcal{\hat{G}}^{\mathcal{S}|\mathcal{T}},\mathcal{G}^\mathcal{T};\mathbf{Y}^{\mathcal{T}}) - I(\mathcal{G}^\mathcal{T};\mathbf{Y}^{\mathcal{T}}) + I(\mathcal{\hat{G}}^{\mathcal{S}|\mathcal{T}},\mathcal{G}^\mathcal{T};\mathbf{Y}^{\mathcal{S}}) - I(\mathcal{G}^\mathcal{T};\mathbf{Y}^{\mathcal{S}}) \\ 
    &= I(\mathbf{Y}^{\mathcal{T}},\mathcal{\hat{G}}^{\mathcal{S}|\mathcal{T}}|\mathcal{G}^\mathcal{T}) + I(\mathbf{Y}^{\mathcal{S}},\mathcal{\hat{G}}^{\mathcal{S}|\mathcal{T}}|\mathcal{G}^\mathcal{T}).
    \end{split}
\end{equation}
Then the Eq~\eqref{eq:lemma_3} can be reformulated as:
\begin{equation}
    \begin{split}
    \label{eq:lemma_7}
    I(\mathcal{\hat{G}}^{\mathcal{S}|\mathcal{T}};\mathbf{Y}|\mathcal{\tilde{G}}^\mathcal{S},\mathcal{G}^\mathcal{T}) \geq 
    I(\mathbf{Y}^{\mathcal{T}},\mathcal{\hat{G}}^{\mathcal{S}|\mathcal{T}}|\mathcal{G}^\mathcal{T}) + I(\mathbf{Y}^{\mathcal{S}},\mathcal{\hat{G}}^{\mathcal{S}|\mathcal{T}}|\mathcal{G}^\mathcal{T}).
    \end{split}
\end{equation}
Then by substituting $I(\mathcal{\hat{G}}^{\mathcal{S}|\mathcal{T}};\mathbf{Y}|\mathcal{\tilde{G}}^\mathcal{S},\mathcal{G}^\mathcal{T})$ in Eq~\eqref{eq:lemma_2} with the above lower bound, we have:
\begin{equation}
\label{eq:lemma_8}
    \begin{split}
        I(\mathcal{\hat{G}}^{\mathcal{S}|\mathcal{T}};\mathcal{\tilde{G}}^\mathcal{S}|\mathcal{G}^\mathcal{T}) & \leqslant 
        -I(\mathbf{Y}^{\mathcal{T}};\mathcal{\hat{G}}^{\mathcal{S}|\mathcal{T}}|\mathcal{G}^{\mathcal{T}})
        -I(\mathbf{Y}^{\mathcal{S}};\mathcal{\hat{G}}^{\mathcal{S}|\mathcal{T}}|\mathcal{G}^{\mathcal{T}}) \\
        & + I(\mathcal{G}^\mathcal{S};\mathcal{\hat{G}}^{\mathcal{S}|\mathcal{T}}|\mathcal{G}^{\mathcal{T}}).
    \end{split}
\end{equation}

\subsection{Proof of Eq~\eqref{eq:opt_pred_1_1} \label{sec:lemma_appendix_2}}
Considering the concept of mutual information and employing variational approximation $p_{\theta}(\mathbf{Y}^{\mathcal{T}}|\mathcal{\hat{G}}^{\mathcal{S}|\mathcal{T}},\mathcal{G}^\mathcal{T})$ to approximate the intractable distribution $p(\mathbf{Y}^{\mathcal{T}}|\mathcal{G}^\mathcal{T})$, we have:
\begin{equation}
\label{eq:lemma_9}
    \begin{split}
    I(\mathbf{Y}^{\mathcal{T}};&\mathcal{\hat{G}}^{\mathcal{S}|\mathcal{T}},\mathcal{G}^\mathcal{T}) = \mathbb{E}_{\mathbf{Y}^{\mathcal{T}},\mathcal{\hat{G}}^{\mathcal{S}|\mathcal{T}},\mathcal{G}^\mathcal{T}}[log\frac{p(\mathbf{Y}^{\mathcal{T}}|\mathcal{\hat{G}}^{\mathcal{S}|\mathcal{T}},\mathcal{G}^\mathcal{T})}{p(\mathbf{Y}^{\mathcal{T}})}] \\ 
    &= \mathbb{E}_{\mathbf{Y}^{\mathcal{T}},\mathcal{\hat{G}}^{\mathcal{S}|\mathcal{T}},\mathcal{G}^\mathcal{T}}[log\frac{p(\mathbf{Y}^{\mathcal{T}}|\mathcal{\hat{G}}^{\mathcal{S}|\mathcal{T}},\mathcal{G}^\mathcal{T})}{p(\mathbf{Y}^{\mathcal{T}})}] \\
    &+ \mathbb{E}_{\mathcal{\hat{G}}^{\mathcal{S}|\mathcal{T}},}[KL(p(\mathbf{Y}^{\mathcal{T}}|\mathcal{\hat{G}}^{\mathcal{S}|\mathcal{T}},\mathcal{G}^\mathcal{T})||p_{\theta}(\mathbf{Y}^{\mathcal{T}}|\mathcal{\hat{G}}^{\mathcal{S}|\mathcal{T}},\mathcal{G}^\mathcal{T}))].
    \end{split}
\end{equation}
From the non-negativity of the KL divergence, we have:
\begin{equation}
\label{eq:lemma_10}
    \begin{split}
        I(\mathbf{Y}^{\mathcal{T}};\mathcal{\hat{G}}^{\mathcal{S}|\mathcal{T}},\mathcal{G}^\mathcal{T}) &\geq \mathbb{E}_{\mathbf{Y}^{\mathcal{T}},\mathcal{\hat{G}}^{\mathcal{S}|\mathcal{T}},\mathcal{G}^\mathcal{T}}[log\frac{p(\mathbf{Y}^{\mathcal{T}}|\mathcal{\hat{G}}^{\mathcal{S}|\mathcal{T}},\mathcal{G}^\mathcal{T})}{p(\mathbf{Y}^{\mathcal{T}})}] \\ 
        &= \mathbb{E}_{\mathbf{Y}^{\mathcal{T}},\mathcal{\hat{G}}^{\mathcal{S}|\mathcal{T}},\mathcal{G}^\mathcal{T}}[logp(\mathbf{Y}^{\mathcal{T}}|\mathcal{\hat{G}}^{\mathcal{S}|\mathcal{T}},\mathcal{G}^\mathcal{T})] + \bm{H}(\mathbf{Y}^{\mathcal{T}}).
    \end{split}
\end{equation}
The same conclusion could be deduced for the term $I(\mathbf{Y}^{\mathcal{S}};\mathcal{\hat{G}}^{\mathcal{S}|\mathcal{T}},\mathcal{G}^\mathcal{T})$.

\subsection{Proof of Eq~\eqref{eq:opt_pred_3_1} \label{sec:lemma_appendix_3}}
After obtaining the compressed behavior subgraph $\mathcal{\hat{G}}^{\mathcal{S}|\mathcal{T}}$ and the corresponding representation matrix $\hat{H}^{\mathcal{S}|\mathcal{T}}$. 
We derive the upper bound of $I(\mathcal{\hat{G}}^{\mathcal{S}|\mathcal{T}};\mathcal{G}^\mathcal{S},\mathcal{G}^\mathcal{T})$ using the variational approximation $p(\hat{H}^{\mathcal{S}|\mathcal{T}})$ of distribution $q(\hat{H}^{\mathcal{S}|\mathcal{T}})$:
\begin{equation}
\label{eq:lemma_11}
    \begin{split}
    I(\mathcal{\hat{G}}^{\mathcal{S}|\mathcal{T}};\mathcal{G}^\mathcal{S},\mathcal{G}^\mathcal{T}) 
    &= \mathbb{E}_{\mathcal{G}^\mathcal{S},\mathcal{G}^\mathcal{T}}[log\frac{p(\mathcal{\hat{G}}^{\mathcal{S}|\mathcal{T}}|,)}{p(\mathcal{\hat{G}}^{\mathcal{S}|\mathcal{T}})}] \\ 
   &- \mathbb{E}_{\mathcal{\hat{G}}^{\mathcal{S}|\mathcal{T}},\mathcal{G}^\mathcal{S},\mathcal{G}^\mathcal{T}}[KL(p(\mathcal{\hat{G}}^{\mathcal{S}|\mathcal{T}})||q(\mathcal{\hat{G}}^{\mathcal{S}|\mathcal{T}}))].
    \end{split}
\end{equation}
Inspired by the non-negativity of KL divergence, and substituting the notation $\mathcal{\hat{G}}^{\mathcal{S}|\mathcal{T}}$ with corresponding representations, we have:
\begin{equation}
\label{eq:lemma_12}
    \begin{split}
    I(\hat{H}^{\mathcal{S}|\mathcal{T}};\mathcal{G}^\mathcal{S},\mathcal{G}^\mathcal{T}) \leq  
    \mathbb{E}_{\mathcal{G}^\mathcal{S},\mathcal{G}^\mathcal{T}}[KL(p(\hat{H}^{\mathcal{S}|\mathcal{T}}|\mathcal{G}^\mathcal{S},\mathcal{G}^\mathcal{T})||q(\hat{H}^{\mathcal{S}|\mathcal{T}}))]
    \end{split}
\end{equation}
Then we assume the aggregated representations $\hat{H}^{\mathcal{S}|\mathcal{T}}$ are completely contaminated with noise $\epsilon \sim N(\mu_{\bm{H}},\sigma^2_{\bm{H}})$, which is sampled from a Gaussian distribution where $\mu_{\bm{H}}$ and $\sigma^2_{\bm{H}}$ are mean and 
variance of $\bm{H}$ which contains information from both the source and target domains. 
Then for each user $j$ in a batch, we have $q(\hat{H}_j^{\mathcal{S}|\mathcal{T}}) = N(\mu_{\bm{H}},\sigma^2_{\bm{H}}).$
Then for $p(\hat{H}^{\mathcal{S}|\mathcal{T}}|\mathcal{G}^\mathcal{S},\mathcal{G}^\mathcal{T})$, we have:
\begin{equation}
\label{eq:lemma_14}
    \begin{split}
        p(\hat{H}^{\mathcal{S}|\mathcal{T}}|\mathcal{G}^\mathcal{S},\mathcal{G}^\mathcal{T}) = 
        N(\sum_{j=1}^{B}\lambda_j \bm{H}_j + \sum_{j=1}^{B}(1-\lambda_j) \mu_{H},\sum_{j=1}^{B}(1-\lambda_j)^2\sigma^{2}_{H}),
    \end{split}
\end{equation}
where $B$ presents the batch size.
Finally, by plugging Eq~\eqref{eq:lemma_11} and Eq~\eqref{eq:lemma_12} into Eq~\eqref{eq:lemma_10}, we have:
\begin{equation}
\label{eq:lemma_15}
    \begin{split}
    \mathop{\min}\
    I(\mathcal{\hat{G}}^{\mathcal{S}|\mathcal{T}};\mathcal{G}^\mathcal{S},\mathcal{G}^\mathcal{T}) & \leq\mathbb{E}_{\mathcal{G}^\mathcal{S},\mathcal{G}^\mathcal{T}}[-\frac{1}{2}logM+\frac{1}{2B}M+\frac{1}{2B}Q^2]\\
    &:=\mathcal{L}_{KL}(\mathcal{\hat{G}}^{\mathcal{S}|\mathcal{T}};\mathcal{G}^\mathcal{S},\mathcal{G}^\mathcal{T}),
    \end{split}
\end{equation}
where $M=\Sigma_{j=1}^{B}(1-\lambda_j)^2$, $Q=\frac{\Sigma_{j=1}^{B}\lambda_j(\mathcal{H}_j-\mu_{\mathcal{H}})}{\sigma_{H}}$. 
$B$ is the bach size.

\clearpage
\bibliographystyle{ACM-Reference-Format}
\balance
\bibliography{sample-base}

\end{document}